\documentclass[a4paper,
               keeplastbox,   
               ]{jacow}
\usepackage{pdfpages,multirow,ragged2e} %
\makeatletter%
	\ifboolexpr{bool{xetex}}
	 {\renewcommand{\Gin@extensions}{.pdf,%
	                    .png,.jpg,.bmp,.pict,.tif,.psd,.mac,.sga,.tga,.gif,%
	                    .eps,.ps,%
	                    }}{}
\makeatother

\ifboolexpr{bool{xetex} or bool{luatex}} 
 {}                                      
 {\usepackage[utf8]{inputenc}}           

\usepackage[USenglish]{babel}

\usepackage{adjustbox}
\usepackage{multirow}
\usepackage{bm}
\usepackage{physics}
\usepackage{wasysym}
\usepackage{float}

\ifboolexpr{bool{jacowbiblatex}}%
 {%
  \addbibresource{jacow-test.bib}
  \addbibresource{biblatex-examples.bib}
 }{}
\begin{document}

\title{The first superconducting final focus quadrupole prototype of the FCC-\lowercase{ee} study}

\author{A. Thabuis, M. Koratzinos, G. Kirby, M. Liebsch, C. Petrone \\ European Organization for Nuclear Research (CERN), Geneva, Switzerland 
  }
	
\maketitle

\begin{abstract}
A single aperture Canted-Cosine-Theta (CCT) quadrupole magnet, made of NbTi superconductors, has been developed for the final focus region of the FCC-ee study. The conductor layout is optimised to mitigate edge effects on one of the two sides of the magnet that typically lead to undesired higher-order multipoles. Experimental results of a prototype, including paraffin wax impregnation and cryogenic temperature measurements, are presented.
The magnet exhibits no training behaviour, surpassing the nominal current during the initial ramp. Field quality is excellent, with higher-order multipoles below $1\times 10^{-4}$ units, consistent with simulations and room temperature tests. These findings confirm the potential of superconducting CCT magnets to offer compact solutions for applications demanding stringent field quality.
\end{abstract}

\section{INTRODUCTION}
The Future Circular Collider (FCC) study plans on building a Higgs and electroweak factory through collisions of positrons and electrons (FCC-ee) \cite{abada_fcc-ee_2019}. Once these particles have been accelerated to their desired energy, strong so-called final focus quadrupole magnets, focus the two individual beams around the so-called interaction region where collisions occur. 

A Canted-Cosine-Theta (CCT) magnet topology has been proposed for the realization of the final-focus quadrupole, following a similar work done at CERN \cite{kirby_hi-lumi_2018}. The conductor layout has been designed such that the edge effects are reduced on one side and kept unchanged on the other side for comparison. 
A prototype of shorter axial length was designed and built in 2019. Its specifications are summarized in Table~\ref{tab:Specifications}. 
The integrated quadrupole component $B_2$ is $2.2$ times less than the full-size final magnet of length $0.7$~m (initially $1.2$~m-long but recently adjusted in the project baseline).
The magnet is wound with $8$ individually insulated superconducting wires. The wires are then connected in series in a joint box positioned at one end of the magnet.

Superconducting magnets need impregnation for better strand-to-strand thermal management and better mechanical integrity. This consists of filling the gaps between windings with an electrically insulating material, improving the overall thermal conductivity of the pack, and better distributing the stress across it. Traditional epoxy-based impregnation often exhibits a training phenomenon, requiring multiple cycles before optimal performance. Recent studies demonstrate that paraffin wax impregnation eliminates this issue \cite{barna_superconducting_2022, daly_improved_2022, araujo_assessment_2024}, solving a bothersome problem of superconducting magnets. Here, we describe a dedicated impregnation setup developed for this purpose.
Magnetic measurements of the demonstrator were initially conducted at room temperature (i.e. warm) \cite{koratzinos_magnetic_2021} and are now presented at cryogenic temperatures (i.e. cold) to assess superconducting magnet performance accurately. The cold tests were conducted in Nov. 2023.

\begin{table}[b]
\centering
\caption{Specification Final-focus Quadrupole Demonstrator}
\label{tab:Specifications}
\begin{adjustbox}{width=\columnwidth}
\def\arraystretch{1.2}
\begin{tabular}{|c|c||c|c|}
\hline
\textbf{Aperture $\diameter$}        & $40$ mm                                                                & \textbf{Inductance}                     & $6.3$ mH        \\ \hline
\textbf{$l_z$}                       & $430$ mm                                                               & \textbf{Joint R @$600$ A}    & $130$ n$\Omega$ \\ \hline
\textbf{Conductor}                   & NbTi                                                                   & \textbf{$I_\text{nom}$ per wire}        & $764.5$ A       \\ \hline
\textbf{Winding pack}                & $8$ wires                                                              & \textbf{$J_e$}                          & $728$ A/mm$^2$  \\ \hline
\textbf{Wire type}                   & \begin{tabular}[c]{@{}c@{}}LHC cable\\ polyimid insul.\end{tabular} & \textbf{peak $B$}                       & $2.95$ T        \\ \hline
\textbf{Wire $\diameter$  w/ insul.} & $0.970$ mm                                                             & \textbf{peak} $\nabla$ $B$                & $100$ T/m       \\ \hline
\textbf{Wire $\diameter$ w/o insul.} & $0.825$ mm                                                             & \textbf{$\int B_2\, dz$ at r=$r^\star$} & $0.315$ T.m     \\ \hline
\textbf{s/c to Cu ratio}             & $0.526$                                                                & \textbf{$T_\text{nom}$}                 & $1.9$ K         \\ \hline
\textbf{$l_{\text{conductor}}$}      & $276$ m                                                                & \textbf{$T_\text{margin}$}              & $2.5$ K         \\ \hline
\end{tabular}
\end{adjustbox}
\end{table}

\begin{figure}[b!]
   \centering
   \includegraphics[width=\columnwidth]{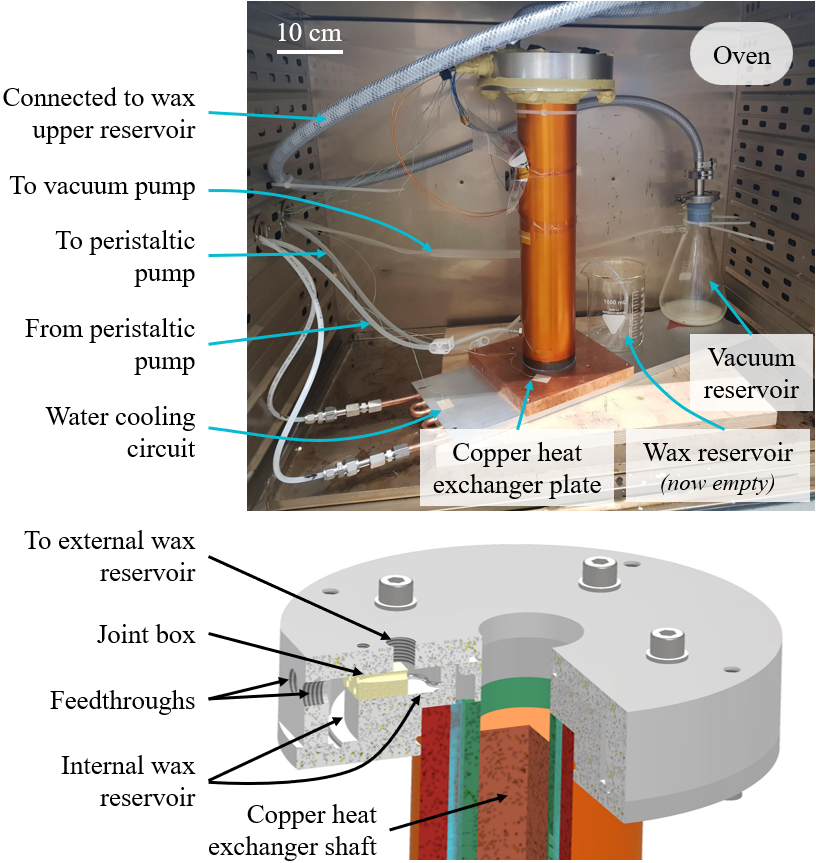}
   \caption{Wax impregnation station inside an oven (top); Cross-section view of the top part of the magnet mounted within the set-up (bottom).}
   \label{fig:waximpregnation}
\end{figure}

\section{WAX IMPREGNATION}
The magnet impregnation process involves injecting hot melted wax into the empty gaps between windings, which then solidifies to ensure mechanical integrity. Paraffin wax undergoes a liquid-solid phase transition at around 55\,°C, contracting by approximately $15\%$ upon solidification. To prevent the formation of trapped empty spaces within the winding pack, temperature gradients are applied across the magnet during progressive cooling, following an approach similar to \cite{barna_superconducting_2022}.

\begin{figure}[t!]
   \centering
   \includegraphics[width=\columnwidth]{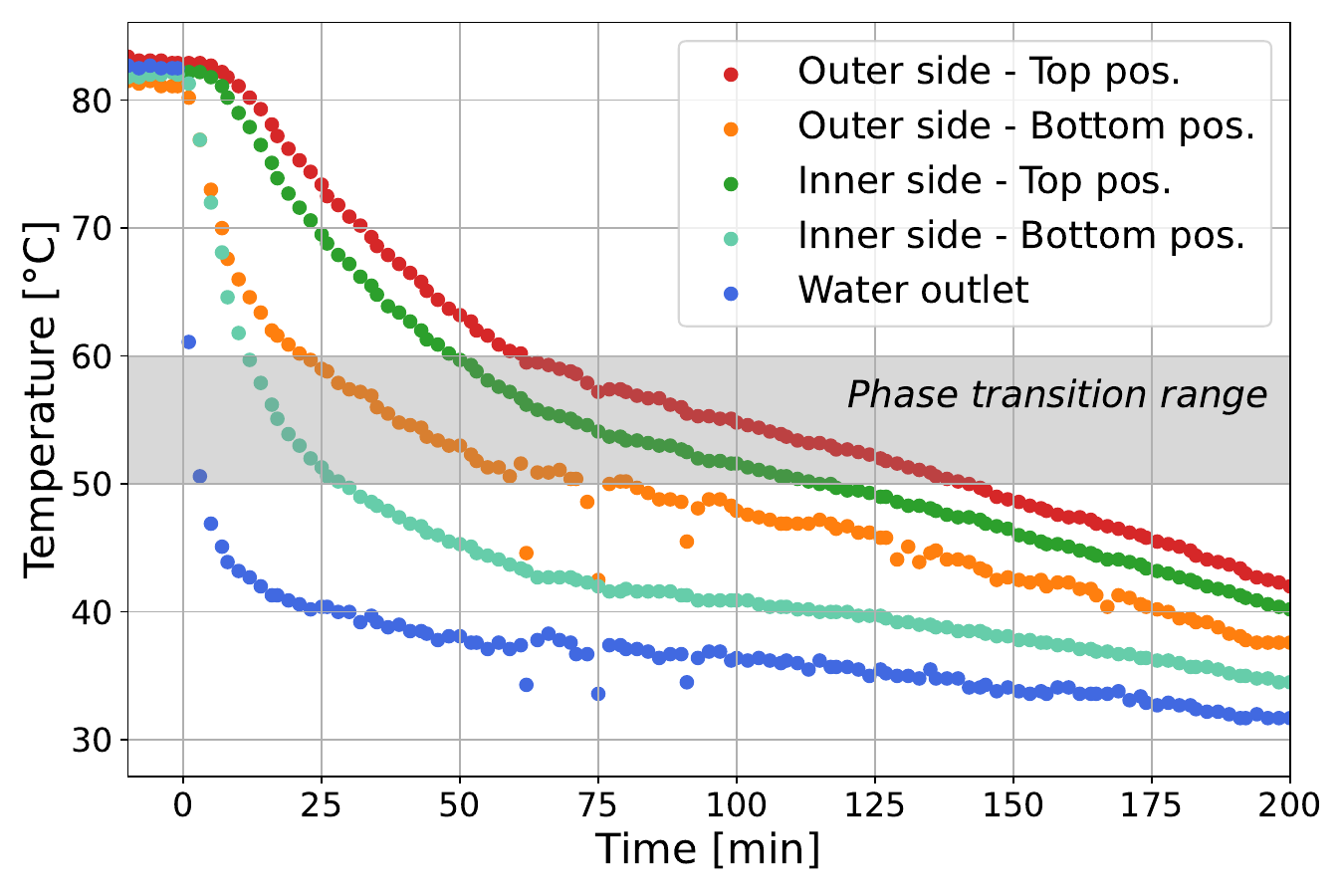}
   \caption{Measured temperature at different positions across the magnet during the wax solidification process.}
   \label{fig:temperature}
   \vspace{-4mm}
\end{figure}

Figure~\ref{fig:waximpregnation} shows the impregnation station and a cross-section of the top part of the magnet. Internal wax reservoirs at the top of the magnet provide additional impregnation volume to compensate for crystallization shrinkage. The system is placed within an oven to achieve the temperature necessary for the wax to be melted. The magnet is air-tight and a weak vacuum is created within the magnet to avoid trapping air bubbles which would degrade the impregnation quality. Wax flow is facilitated by a peristaltic pump, filling the magnet from the bottom.
A copper shaft is inserted inside the magnet and connected to a copper plate lying at the bottom of it. This induces two temperature gradients across the magnet: one from the inner layer to the outer layer, and another from the top to the bottom part of the magnet. A water cooling circuit is added below the magnet enhancing temperature control of the heat exchanger and accelerating wax cooling.

Figure~\ref{fig:temperature} shows the temperature profiles measured at various positions across the magnet and downsampled by averaging over ten points. An average 3\,°C and 5\,°C temperature difference was observed between the inner side (in contact with the heat exchanger shaft) and outer side (in contact with the magnet housing); and between the top and bottom part of the magnet respectively.

\section{TESTING CAMPAIGN AT COLD}
\begin{figure}[b!]
	\centering
	\includegraphics[width=.95\columnwidth]{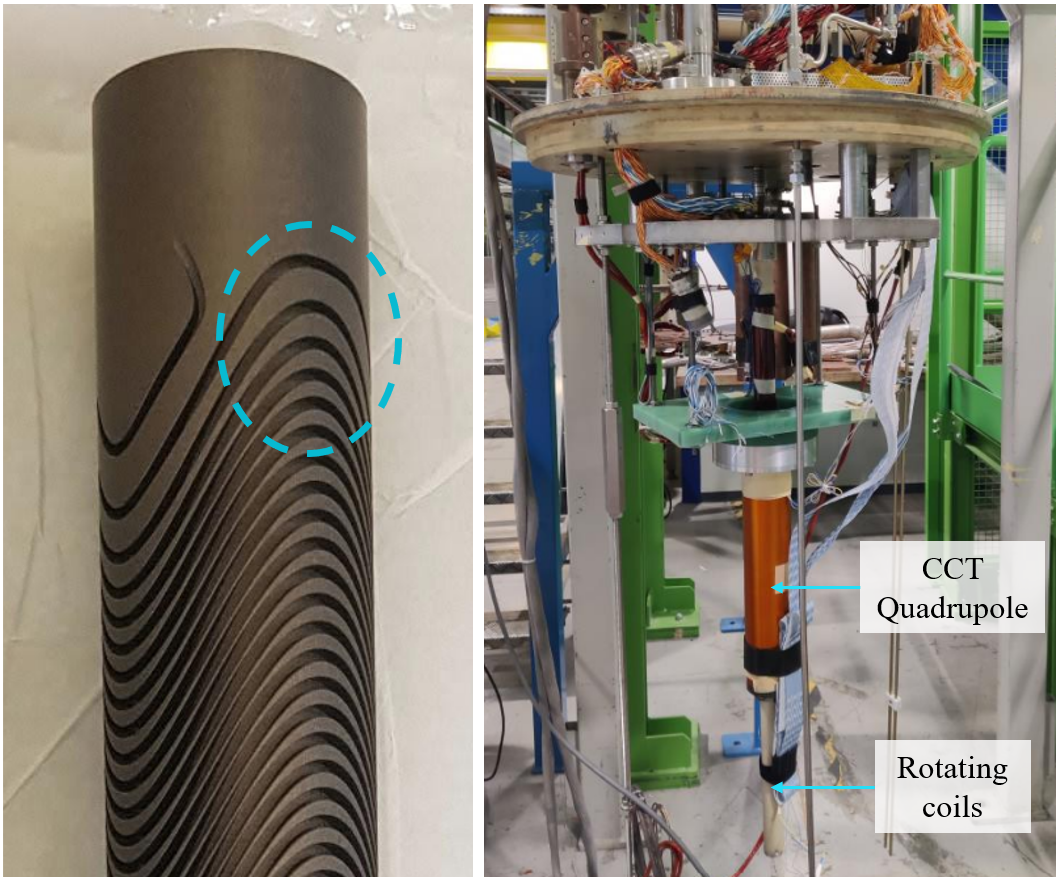}
	\caption{(Left) Inner former of the CCT magnet. Note the deviation of the first two turns from a pure sinusoidal shape \cite{koratzinos_method_2018}. (right) Open cryostat set-up for cold measurements.}
	\label{fig:CCTformer_SetUp}
\end{figure}

Once impregnated, the magnet was tested at cryogenic temperatures for characterization. The prototype was placed inside a cryostat with rotating coils, as depicted on the right side of Fig.~\ref{fig:CCTformer_SetUp}. The cryostat is filled with helium and cooled to $1.9$~K and later to $4.5$~K. The test took place in November 2023 at the Siegtag cryostat of the SM18 cryogenic facility of CERN.

\begin{figure}[b!]
   \centering
   \includegraphics[width=\columnwidth]{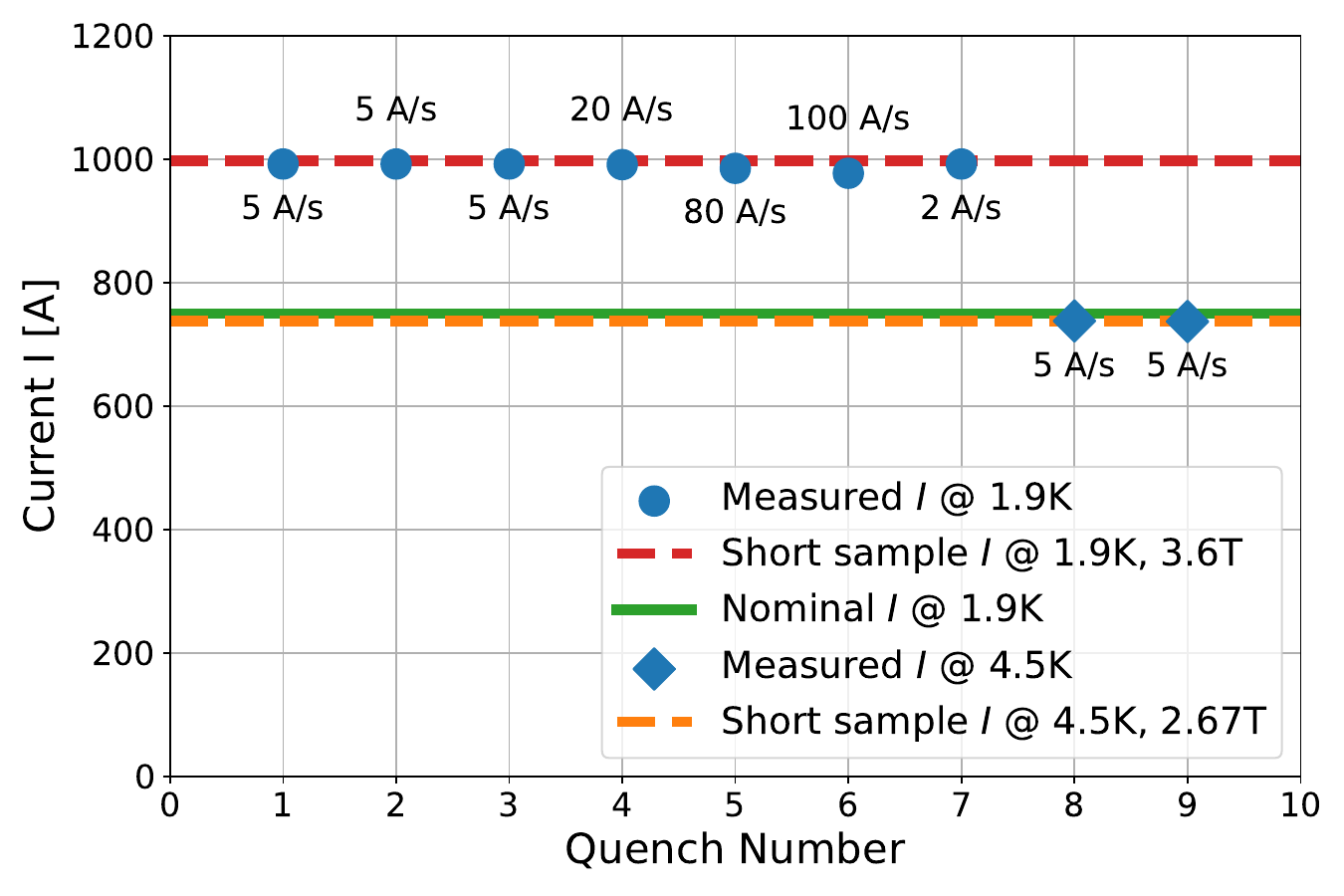}
   \caption{Quenching campaign of the magnet with various current ramping rates.}
   \label{fig:training_quench}
\end{figure}
 
\subsection{Critical Current Density}
Multiple taps within the joint box monitor voltage between wires to detect quenches and protect the magnet by discharging energy into a dump resistor. A $50$ $\Omega$ crowbar connected to a $75$ $\Omega$ dump resistor is employed for energy extraction. A $100$ mV threshold with $10$ ms validation time is used for quench detection. Inductive voltages are canceled by comparing the first half of the magnet with the second half, and quarters of the magnets are also compared for redundancy and in the case of symmetric quenches.

Various ramping tests determine the critical current of the magnet, as illustrated in Fig.~\ref{fig:training_quench}. Notably, no training is observed. The short-sample critical current already measured on previous occasions is reached on the first attempt. A margin exceeding $20\%$ at $1.9$ K is validated for the nominal operating current.
Additional tests have been performed at $4.5$ K for demonstration, also reaching the short sample limit associated with this operating temperature.

\newcommand{\specialtabular}[1]{%
  \renewcommand{\arraystretch}{1}%
  \begin{tabular}[c]{@{}c@{}}#1\end{tabular}%
}

\renewcommand{\arraystretch}{1.5}
\begin{table}[b!]
\centering
\caption{Comparison of the field quality between the warm and cold ($1.9$ K) measurements of the CCT quadrupole powered with a current of $5$ A and $765$ A respectively. The normal $b_n$ and skew $a_n$ components of different higher order multipoles $n$ are evaluated at a $R_\text{ref}=10$ mm. They are expressed in $10^{-4}$ units.}
\label{tab:fieldquality}
\begin{adjustbox}{width=0.5\textwidth}
\begin{tabular}{|c|c||ccc||ccc|}
\hline \cline{1-8}
\multicolumn{1}{|c|}{} & \multicolumn{1}{c||}{} & \multicolumn{3}{c||}{$\bm{b_n}$}   & \multicolumn{3}{c|}{$\bm{a_n}$}   \\ \hline  \cline{1-8}
$\bm{n}$                 & \textbf{Case}       & \multicolumn{1}{c|}{\textbf{Center}} & \multicolumn{1}{c|}{\specialtabular{\textbf{Non} \\ \textbf{Corrected} \\ \textbf{Side}}} & {\specialtabular{\textbf{Corrected} \\ \textbf{Side}}} & \multicolumn{1}{c|}{\textbf{Center}} & \multicolumn{1}{c|}{{\specialtabular{\textbf{Non} \\ \textbf{Corrected} \\ \textbf{Side}}}} & {\specialtabular{\textbf{Corrected} \\ \textbf{Side}}} \\ \hline \cline{1-8} 
\multirow{2}{*}{\textbf{3}}     & Warm                  
        & \multicolumn{1}{c|}{$-0.22$}    & \multicolumn{1}{c|}{$-35.98$}        & \multicolumn{1}{c||}{$1.83$}   
        & \multicolumn{1}{c|}{$0.32$}    & \multicolumn{1}{c|}{$-78.07$}        & \multicolumn{1}{c|}{$-2.24$}       \\ 
                       & Cold          
        & \multicolumn{1}{c|}{$0.32$}    & \multicolumn{1}{c|}{$-32.04$}        & \multicolumn{1}{c||}{$0.71$}   
        & \multicolumn{1}{c|}{$0.11$}    & \multicolumn{1}{c|}{$-57.36$}        & \multicolumn{1}{c|}{$-0.78$}       \\ \hline \cline{1-8}
\multirow{2}{*}{\textbf{4}}     & Warm                  
        & \multicolumn{1}{c|}{$0.53$}    & \multicolumn{1}{c|}{$11.80$}        & \multicolumn{1}{c||}{$3.38$}   
        & \multicolumn{1}{c|}{$0.54$}    & \multicolumn{1}{c|}{$-46.56$}        & \multicolumn{1}{c|}{$-1.15$}       \\ 
                       & Cold          
        & \multicolumn{1}{c|}{$0.52$}    & \multicolumn{1}{c|}{$13.81$}        & \multicolumn{1}{c||}{$3.36$}   
        & \multicolumn{1}{c|}{$0.79$}    & \multicolumn{1}{c|}{$-37.39$}        & \multicolumn{1}{c|}{$-1.44$}       \\ \hline \cline{1-8}
\multirow{2}{*}{\textbf{5}}     & Warm                  
        & \multicolumn{1}{c|}{$-0.16$}    & \multicolumn{1}{c|}{$-8.66$}        & \multicolumn{1}{c||}{$-0.30$}   
        & \multicolumn{1}{c|}{$-0.03$}    & \multicolumn{1}{c|}{$-5.83$}        & \multicolumn{1}{c|}{$0.02$}       \\ 
                       & Cold          
        & \multicolumn{1}{c|}{$-0.18$}    & \multicolumn{1}{c|}{$-7.31$}        & \multicolumn{1}{c||}{$-0.30$}   
        & \multicolumn{1}{c|}{$-0.06$}    & \multicolumn{1}{c|}{$-4.13$}        & \multicolumn{1}{c|}{$0.18$}       \\ \hline \cline{1-8}
\multirow{2}{*}{\textbf{6}}     & Warm                  
        & \multicolumn{1}{c|}{$0.64$}    & \multicolumn{1}{c|}{$4.44$}        & \multicolumn{1}{c||}{$1.02$}   
        & \multicolumn{1}{c|}{$-0.12$}    & \multicolumn{1}{c|}{$1.74$}        & \multicolumn{1}{c|}{$-0.11$}       \\ 
                       & Cold          
        & \multicolumn{1}{c|}{$0.79$}    & \multicolumn{1}{c|}{$4.11$}        & \multicolumn{1}{c||}{$1.22$}   
        & \multicolumn{1}{c|}{$-0.07$}    & \multicolumn{1}{c|}{$1.80$}        & \multicolumn{1}{c|}{$-0.12$}       \\ \hline \cline{1-8}
\multirow{2}{*}{\textbf{7}}     & Warm                  
        & \multicolumn{1}{c|}{$0.03$}    & \multicolumn{1}{c|}{$-1.62$}        & \multicolumn{1}{c||}{$0.00$}   
        & \multicolumn{1}{c|}{$-0.00$}    & \multicolumn{1}{c|}{$-0.79$}        & \multicolumn{1}{c|}{$-0.01$}       \\ 
                       & Cold          
        & \multicolumn{1}{c|}{$0.01$}    & \multicolumn{1}{c|}{$-1.47$}        & \multicolumn{1}{c||}{$-0.03$}   
        & \multicolumn{1}{c|}{$-0.01$}    & \multicolumn{1}{c|}{$-0.91$}        & \multicolumn{1}{c|}{$0.01$}       \\ \hline \cline{1-8}
\multirow{2}{*}{\textbf{8}}     & Warm                  
        & \multicolumn{1}{c|}{$0.00$}    & \multicolumn{1}{c|}{$0.65$}        & \multicolumn{1}{c||}{$-0.05$}   
        & \multicolumn{1}{c|}{$-0.03$}    & \multicolumn{1}{c|}{$0.27$}        & \multicolumn{1}{c|}{$-0.02$}       \\ 
                       & Cold          
       & \multicolumn{1}{c|}{$0.00$}    & \multicolumn{1}{c|}{$0.65$}        & \multicolumn{1}{c||}{$-0.01$}   
        & \multicolumn{1}{c|}{$-0.02$}    & \multicolumn{1}{c|}{$0.35$}        & \multicolumn{1}{c|}{$-0.04$}       \\ \hline \cline{1-8}
\end{tabular}
\end{adjustbox}
\end{table}

\subsection{Field Quality Measurements}
The field quality is determined by the normalized coefficients resulting from the Fourier analysis of the flux density distribution across the magnet cross-section \cite[Chapter~6]{russenschuck_field_2011}. 
These coefficients are determined by inductive passive PCB coils rotating around the axis inside the magnet. 
Three coils cover the entire axial length: one at the center for nominal magnetic strength characterization, and two others to investigate edge effects. A distinctive conductor layout on one side corrects undesired higher-order multipole content. This is illustrated in Fig.~\ref{fig:CCTformer_SetUp} with the grooves hosting the conductor diverging from the usual sinusoidal profile that leads to undesired edge effects. 
The correction is confirmed by comparing coefficients from the coils positioned on the corrected and non-corrected sides as shown in TABLE~\ref{tab:fieldquality}. 
%
The field quality analysis results, although worse than the warm tests, are still below $1$ unit in the central region. We have not repeated the extra step done at warm which reduced multipole errors and consisted of rotating the magnet by about $45^\circ$ with respect to the measuring system (before repeating the measurements) \cite{koratzinos_magnetic_2021}. 

The magnitude of undesired higher-order harmonics is significantly reduced on the corrected side compared to the non-corrected one, reaching values below $5$ units (the normalization uses the quadrupole field of that side and not of the whole magnet). 
One should note that when evaluating the overall performance of the magnet, the total length would be considered, resulting in a larger normalizing quadrupole component which mainly comes from the central part. This dilutes the contribution of the edges to the field quality of the prototype, and even more for the full-scale magnet.

This validates the possibility of tuning the conductor layout of CCT magnets in the edge to remove undesired higher-order multipole content. This approach can also get rid of crosstalk from the adjacent magnet in a twin aperture arrangement as envisaged in FCC-ee.

\section{CONCLUSION}
This work presents the first prototype built for the final focus quadrupole of the FCC-ee study. A CCT superconducting magnet impregnated with paraffin wax is built and tested. The magnet exhibits no training behavior and reaches the short-sample critical current ($1000$ A) for various current rates up to $100$ A/s. The field quality is excellent. Additionally, this prototype validates the possibility of tuning the conductor layout of a CCT magnet to greatly reduce edge effects for improved field quality. Future work will present a similar approach for crosstalk compensation of a twin-aperture arrangement. This will be demonstrated with a CCT sextupole wound with High-Temperature-Superconducting (HTS) tapes.

\section{ACKNOWLEDGMENTS}
We would like to acknowledge the help of Michael Daly, Colin Müller, and Jaap Kosse from the Paul Scherrer Institute (PSI), as well as Daniel Barna from the Wigner Research Centre for Physics for their support with the impregnation station. Additionally, many thanks to the team at SM18 Arnaud Devred, Gerard Willering, Jerome Feuvrier, and Franco Julio Mangiarotti for their help with the cold tests.
Finally, we would like to thank Austin Ball, Günther Dissertori, Markus Klute, and Maf Alidra for their invaluable help during various stages of the project.

%
%
\ifboolexpr{bool{jacowbiblatex}}%
	{\printbibliography}%
	{%
	
 
} 

\end{document}